# Variability of structural and electronic properties of bulk and monolayer Si$_2$Te$_3$

Y. S. Puzyrev[1], X. Shen[1,2], C. Combs[1], and S. T. Pantelides[1,3]

[1]Department of Physics and Astronomy, Vanderbilt University, Nashville, TN 37235

[2]Department of Physics and Materials Science, University of Memphis, Memphis, TN 38152

[3]Department of Electrical Engineering and Computer Science, Vanderbilt University, Nashville, TN 37235

Since the emergence of monolayer graphene as a promising two-dimensional material, many other monolayer and few-layer materials have been investigated extensively. An experimental study of few-layer Si$_2$Te$_3$ was recently reported, showing that the material has diverse properties for potential applications in Si-based devices ranging from fully integrated thermoelectrics to optoelectronics to chemical sensors. This material has a unique layered structure: it has a hexagonal closed-packed Te sublattice, with Si dimers occupying octahedral intercalation sites. Here we report a theoretical study of this material in both bulk and monolayer form, unveiling a fascinating array of diverse properties arising from reorientations of the silicon dimers between planes of Te atoms. The lattice constant varies up to 5% and the band gap varies up to 40% depending on dimer orientations. The monolayer band gap is 0.4 eV larger than the bulk-phase value for the lowest-energy configuration of Si dimers. These properties are, in principle, controllable by temperature and strain, making Si$_2$T$_3$ a promising candidate material for nanoscale mechanical, optical, and memristive devices.

The discovery and characterization of new two-dimensional materials, sparked by graphene's remarkable properties, holds a promise for emerging technologies, especially, for nanoscale mechanical and optical devices. Two-dimensional materials create opportunities for surface functionalization and tailoring mechanical, electronic and optical properties, exemplified by graphene[1–3], dichalcogenides such as $MoS_2$,[4,5] and black phosphorous.[6–10]

Isolation of few-layer $Si_2Te_3$, as few as five layers, was demonstrated in a recent publication[11]. Compared with other layered materials, the crystal structure of $Si_2Te_3$ is quite complex. In $Si_2Te_3$, the Te atoms form a hexagonal close packed sublattice, while Si atoms pair into Si-Si dimers and fill 2/3 of the possible interstitial sites (Figure 1a). There are four possible orientations of the Si-Si dimers, three of them are in-plane (horizontal) directions that are at 60 degrees to each other (Figure 1b-d), and one out-of-plane (vertical) direction, perpendicular to the 2D plane (Figure 1e). X-ray and electron diffraction data on bulk $Si_2Te_3$ suggest that a quarter of the dimers are vertical, while the other three quarters of the dimers are oriented horizontally with a random choice of 0°, 30°, or 60° angle. [12,13]

Recent progress in developing techniques of isolating single-layer structures, e.g., laser thinning of $MoS_2$, recently used by Cho et al. to thin down $MoTe_2$,[14] shows promise for making single-layer $Si_2Te_3$. Experimentally, it has been observed that the $Si_2Te_3$ nanostructures exhibit a reversible change of color from red to black upon heating to 210 °C,[11] suggesting a reduction of the band gap.

In this Letter, we report a theoretical study of bulk and monolayer $Si_2Te_3$ using density functional theory (DFT) which examines the structures, as observed in x-ray and electron diffraction experiments, and identifies the equilibrium atomic structure of $Si_2Te_3$. We also demonstrate a large variation of band-gap and lattice constants of the material for both a monolayer and bulk material. We show that the key to explaining the variation of the mechanical and electronic properties is the orientation of Si dimers incorporated in Te layers. Calculations of the Si dimers at elevated temperature suggest the possibility of controlling the change of dimer orientation, which can lead to a change in lattice constant and band gap.

Current understanding of the $Si_2Te_3$ crystal structure is based on x-ray and electron diffraction experiments on bulk crystals.[12,13] Both methods conclude that the $Si_2Te_3$ has a layered structure, where each

Te-Si-Te layer contains two planes of Te atoms that are hexagonally close packed, as shown in Fig. 1a. Si atoms form Si-Si dimers and occupy 2/3 of the octahedron sites between the hcp Te atoms. The repeating unit along the vertical direction of the HCP lattice contains two Te-Si-Te layers, in which the Si-Si dimers occupy different sets of the 2/3 of the octahedron sites.

Results of x-ray and electron diffraction experiments indicate that three quarters of dimers orient along three equivalent $[1\bar{1}00]$ directions of the HCP lattice, shown in Fig. 1b-d. The remaining quarter of the Si-Si dimers are parallel to the [0001] direction of the HCP lattice, Fig. 1e. It is also found that the $Si_2Te_3$ undergoes a phase transition at 350 °C, where the Si-Si dimers dissociate and the resulting individual Si atoms migrate to the tetrahedral interstitials sites between the Te atoms.[12]

We have performed calculations using DFT with the local-density approximation for the exchange-correlation potential, PAW potentials,[15] and plane-wave basis as implemented in the VASP code.[16,17] Supercells including 20, 40 and 80 atoms with different Si dimer orientation, as observed in x-ray and electron diffraction experiments[12,13], were used in the calculations. A 10×10×10 k-point grid was employed for the Brillouin zone integrations for structural optimization and, subsequently, the band structure was calculated for high-symmetry directions. Atomic positions were relaxed until the configuration energy difference was less than $10^{-3}$ eV.

First-principles molecular dynamics (MD) calculations were performed to test the stability of Si dimers at elevated temperature up to 1000 K. The calculations ran for 1 ps using an 80-atom supercell and resulted in a Si dimer flip. To extend the simulation time to a picosecond, the MD calculations were performed using only the Γ point in reciprocal space.

In order to assess the energy barrier that must be overcome to cause dimer reorientation at finite temperature, we performed a nudged-elastic-band calculation.[18] The initial and final configurations corresponding to an MD-simulated dimer reorientation were used to determine the reorientation path and energy barrier, as shown in Fig. 2.

The results of DFT calculations show that, in both bulk and monolayer configuration, $Si_2Te_3$ has the lowest energy when all the Si dimers are aligned horizontally in the same direction, as shown in Fig. 1b-d and Fig. 3b. Meanwhile, rotating the direction of a single Si dimer horizontally in a bulk material, as shown in Fig. 2a, raises the total energy of the system by 17 meV. Thus, horizontally misaligned dimers can exist at room temperature. Flipping the direction of a Si-Si bond from the horizontal to the vertical direction raises the total energy of the system by 130 meV. Vertical Si dimers, which are observed in the x-ray and electron diffraction experiments[11], are the results of flipping horizontal dimers under thermal or other excitations, caused by x-ray or electron beams.

Molecular dynamics simulations show that in the lowest energy configuration of $Si_2Te_3$, Si dimers change orientation, as shown in Fig. 2a, at 700 K within a short period of simulation time, namely ~1 ps. This reorientation has energy barrier of 1eV in bulk $Si_2Te_3$, calculated using NEB, making this transition plausible at room temperature, and likely at elevated temperatures. Thus an equilibrium structure of $Si_2Te_3$ at macroscopic scale may consist of regions with various orientations of Si dimers.

We also found that the band structure changes upon the flipping of Si-Si dimers. For example, theory predicts that in the bulk material, the band gap of the $Si_2Te_3$ changes from indirect, from H to A points in reciprocal space, shown in Fig. 2(a, b), to direct at point A and decreases by 0.4 eV when the dimers take all possible orientations (Fig. 2c). The monoloyer $Si_2Te_3$ also exhibits a change in the band gap value. However, the monolayer band gap remains indirect and reduces from a 1.8 eV value from $\Gamma$ to K for a configuration where all pairs are oriented in the same horizontal direction, Fig. 4a, to 1.6 eV from M to K, where a quarter of pairs are oriented vertically.

In both bulk and monolayer $Si_2Te_3$, the lattice constant depends on the orientation of the dimers, with the larger value for the lowest energy configuration of all dimers oriented horizontally in the same direction. However, a configuration, where a quarter of the dimers are vertically oriented has a lattice constant reduced by 5% and 4% for bulk and monolayer cases respectively. As we already noted, reorientation of a single Si dimer from horizontal to vertical position only raises the total energy of the system by 0.13 eV with a barrier of 1 eV. Thus, vertically positioned dimers are ubiquitous at and above room temperature. The regions containing vertical dimers can be viewed as regions containing defects, due to local 'defect' levels, corresponding to the fact that $Si_2Te_3$ with vertically oriented dimers has a lower band gap. In a device, electrons injected from the electrodes can be trapped in these defect levels, creating regions with high concentration of charged 'defects'. Electric-field-driven motion of charged defects is likely to cause resistance change in the material by altering the electrostatic potentials, similar to many transition metal oxides where the motion of defects results in a resistance hysteresis and memristive switching.

The lowest energy state of $Si_2Te_3$ is three-fold degenerate and, given a small energy for dimer reorientation, the material would contain all possible orientations at finite temperature. We expect that at a finite temperature, $Si_2Te_3$ can be composed of domains, where the dimers are aligned, and the domain boundaries, described as the adjacent regions with different dimer orientations. The domain boundaries can be viewed as collective structural 'excitation', which can act as a host to excitons, localized at the domain boundaries, where band structure is preferable for exciton generation.

Large changes in the lattice constant and the concomitant band structure changes caused by mere re-orientation of Si dimers suggests that extensive tuning of structural and electronic properties can in principle be achieved by applying strain in one, two, or three dimensions. Electronic, optical, and sensor applications may be possible. In Ref. 11, it was demonstrated that Li+ and Mg+ ions can be intercalated in $Si_2Te_3$, opening the possibility of battery applications.

Acknowledgement

This work was supported by National Science Foundation grant number DMR-1508433. C. Combs is an undergraduate at University of Tennessee at Knoxville and was supported during the summer of 2015 by NSF REU grant EPS-1004083. The calculations were performed using the Extreme Science and Engineering Discovery Environment (XSEDE), which is supported by grant number ACI-1053575 and the High Performance Computing Facilities at University of Memphis.

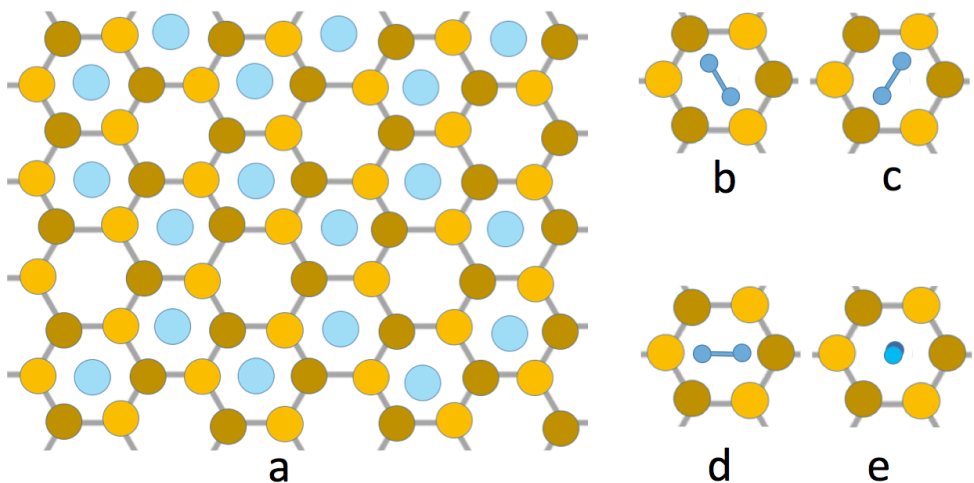

Figure 1. (a) Top view of hcp Te layers, upper and lower layers of Te are shown in light and dark tan color respectively. The light blue sites are occupied by Si-Si dimers between the two Te planes. (b-e) Four possible orientations of the Si-Si dimers, shown in blue. (b-d) In-plane orientations: Si dimer is parallel to one of the three $[1\bar{1}00]$ directions. (e) Out-of-plane orientation.

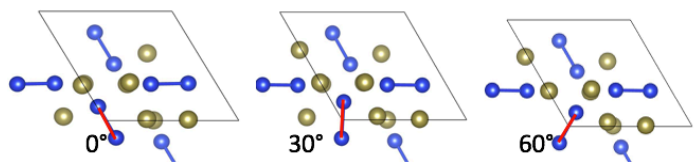
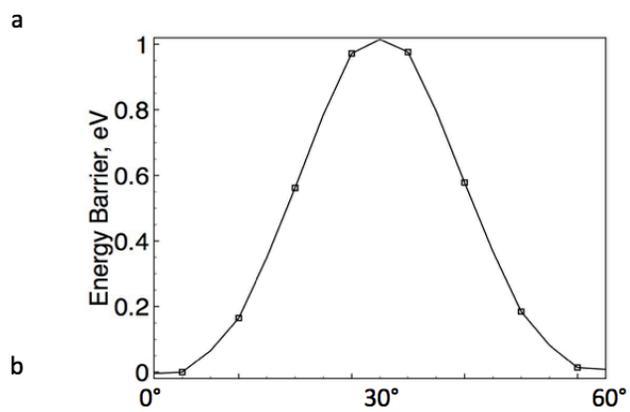

Figure 2. (a) Si dimer reorientation pathway and (b) energy barrier of the horizontal rotation of a Si dimer (marked in red).

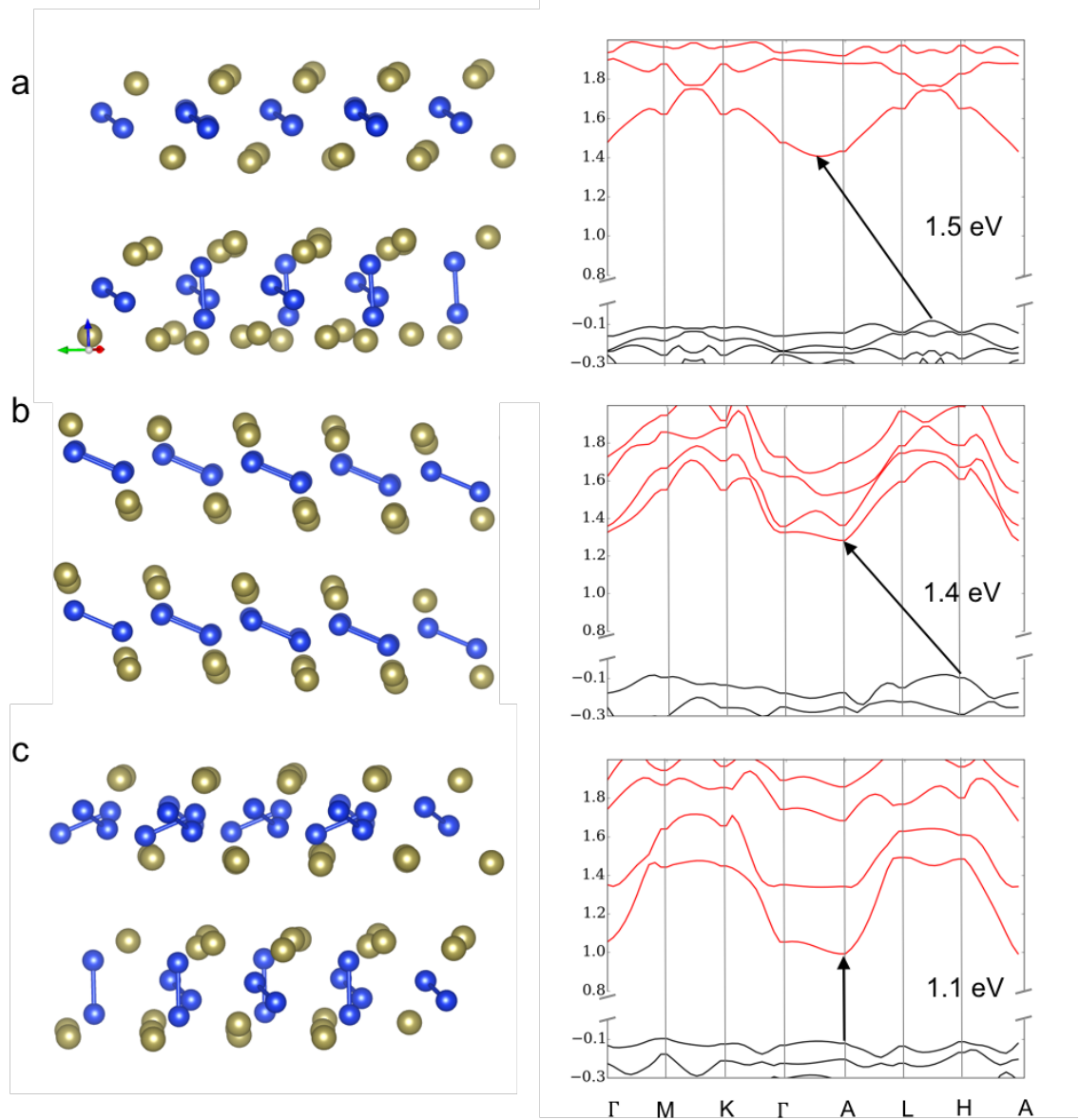

Figure 3. Low-energy configurations of dimer orientations in in bulk $Si_2Te_3$ with corresponding band structure changing its character from a) and b) indirect to c) direct and varying from 1.5 to 1.1 eV.

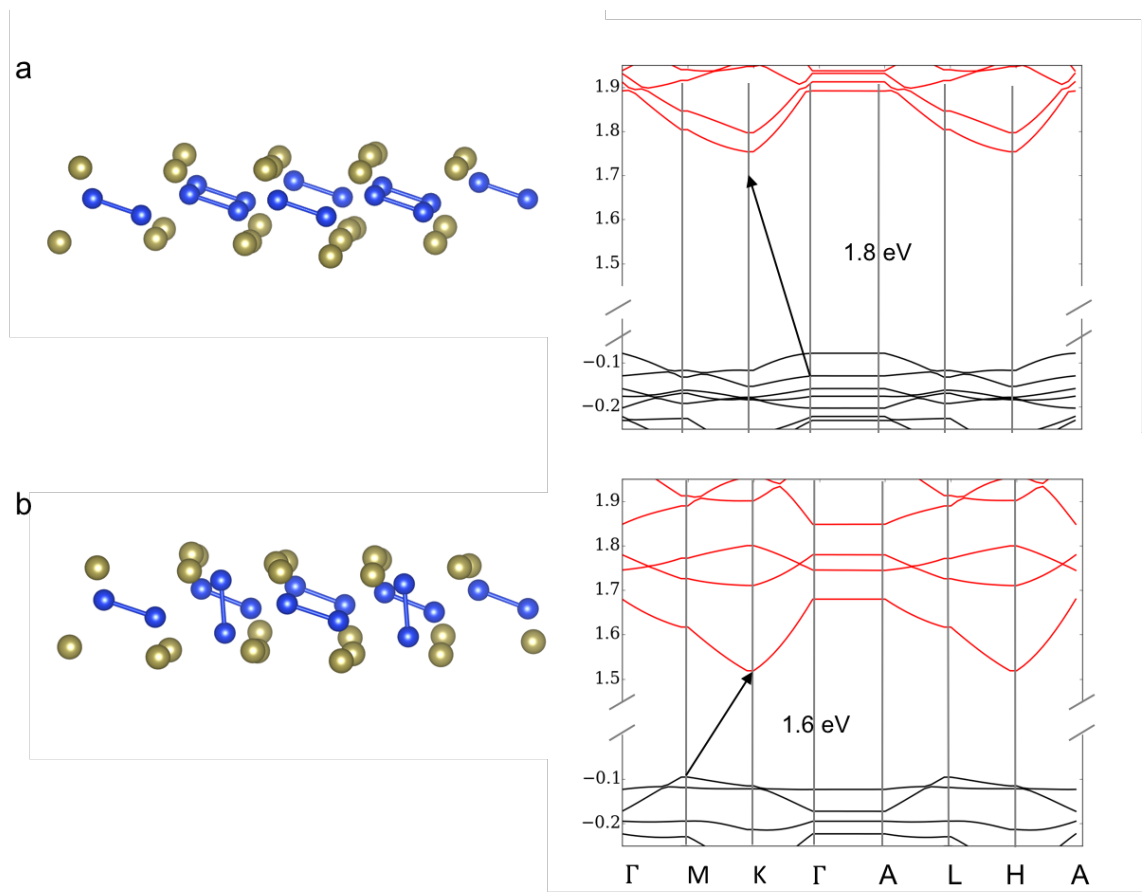

Figure 4. Two-dimensional structure, with a) all dimers oriented parallel to Te layers and b) quarter of dimers are oriented vertically.